# Self-Compensation of Conduction in $Cd_{0.95}Zn_{0.05}Te$:Cl Crystals in a Wide Range of Cd Vapor Pressures

O. A. Matveev\*, A. I. Terent'ev\*, N. K. Zelenina\*, V. N. Gus'kov\*\*, V. E. Sedov\*, A. A. Tomasov\*, and V. P. Karpenko\*

*Ioffe Physicotechnical Institute, Russian Academy of Sciences, St. Petersburg, 194021 Russia*
*e-mail: Oleg.Matveev@mail.ioffe.ru*
**Kurnakov Institute of General and Inorganic Chemistry, Russian Academy of Sciences, Moscow, 117907 Russia*

**Abstract**—The process of self-compensation in $Cd_{0.95}Zn_{0.05}Te$:Cl solid-solution crystals has been studied by annealing single crystals under a controlled Cd vapor pressure, with subsequent measurements of the Hall effect, photoluminescence, carrier lifetime and mobility, and photocurrent memory in the annealed crystals. By means of this annealing, conditions of thermal treatment that make it possible to fabricate low-conductivity samples with a low carrier density, $10^7$-$10^{11}$ cm$^{-3}$, are denned. In these samples, a $p - n$ conduction inversion is observed at a higher free-carrier density ($n, p \sim 10^9$ cm$^{-3}$) and the dependence of the electron density on the Cd vapor pressure exhibits a more gentle slope than in the case of CdTe:Cl crystals. The obtained data are discussed in terms of a self-compensation model in which intrinsic point defects act as acceptors with deep levels. This level is attributed to a Zn vacancy, which remains active at high Cd pressure.

The synthesis of semi-insulating $Cd_{1-x}Zn_xTe$:Cl crystals is being widely studied at present [1, 2]. The low conductivity of these crystals is accounted for by self-compensation of charged point defects. The two mechanisms most frequently used to explain the self-compensation in a semiconductor doped with a donor impurity are the generation of oppositely charged intrinsic point defects [3-8] and the crystal lattice relaxation, which results in the formation of *DX* centers [9-11]. A high degree of self-compensation in a crystal is reached if it undergoes a slow post-growth cooling, when the interaction of charged point defects occurs [3, 4, 12].

We studied the self-compensation in CdTe:Cl by annealing single crystal samples under a controlled Cd vapor pressure $P_{Cd}$, with subsequent measurement of the Hall effect [13, 14]. The obtained data allowed us to determine the thermal conditions necessary for obtain-ing samples with low conductivity, $\sim 10^{-10}$ $\Omega^{-1}$ cm$^{-1}$, and a low density of free carriers, $10^7$-$10^8$ cm$^{-3}$. The nature of the *p-n* conduction inversion as a function of $P_{cd}$ at low carrier densities, $10^7$-$10^8$ cm$^{-3}$, was established. On the Basis on these results, we determined the conditions necessary for controlling self-compensation during the post-growth annealing of a CdTe:Cl ingot.

Semi-insulating crystals with *p*-type conduction were reproducibly obtained via sample annealing and the growth of a CdTe:Cl ingot. Semi-insulating n-type crystals were obtained considerably less frequently, and, often, they exhibited inhomogeneous physical characteristics. However, even the earliest experiments on the growth of $Cd_{1-x}Zn_xTe$:Cl ($x = 0.0002$-$0.1$) demonstrated a reliable production of semi-insulating *n*-type crystals [15]. It is precisely this kind of n-type semi-insulating crystal that is demanded for the fabrication of X-ray detectors for computer tomography [15].

In this report, we present the results of investigation of single-crystal $Cd_{0.95}Zn_{0.05}Te$:Cl samples annealed under a controlled Cd vapor pressure.

In preparation for annealing, samples were cut from a $Cd_{0.95}Zn_{0.05}Te$:Cl ingot grown by horizontal unidirectional crystallization under controlled $P_{Cd}$ [15]. The growth conditions (the melt overheating, the deviation from stoichiometry, and the growth rate) were the same as in the case of CdTe:Cl [16]. The Cl concentration in the ingot was defined by the $CdCl_2$ charge in the melt, $2 \times 10^{-18}$ cm$^{-3}$ (which was the same for CdTe:Cl). Before annealing, the as-grown $Cd_{0.95}Zn_{0.05}Te$:Cl samples demonstrated *p*-type conduction with a hole density $p = 10^8$-$10^9$ cm$^{-3}$. The method used for annealing was described in detail for CdTe:Cl [13]. The samples were annealed in a three-zone furnace in a quartz ampule. The annealing temperature was $t_{cr}$ = 900°C, the temperature of the filling material was $t_p$ - 905°C, and the temperature of metallic Cd $t_{cd}$ was defined in each experiment by the desired partial pressure of Cd. The filling material was prepared from the same crystal as the sample to be annealed, which meant that the sample composition remained unchanged during the annealing.

The density of carriers was determined from Hall measurements of $Cd_{0.95}Zn_{0.05}Te:Cl$ samples annealed under a selected partial pressure of Cd vapor. Figure 1a (points 2, 2') shows the dependence of the hole ($p$) and electron ($n$) density in these samples on $P_{Cd}$.

Points 1 and 1' in Fig. 1a show similar results for CdTe:Cl. At low $P_{Cd}$, both CdTe:Cl and $Cd_{095}Zn_{005}Te:Cl$ demonstrate p-type conduction. At intermediate $P_{Cd}$, a $p \longrightarrow n$ inversion of the conduction is observed.

At a high partial pressure of Cd, the crystals switch to $n$-type conduction. It is necessary to note that the conduction inversion for the CdTe:Cl crystals is observed within a very narrow range of $P_{Cd}$ values and at low carrier densities, $p(n) = 10^7$-$10^8$ cm$^{-3}$.

In the $Cd_{0.95}Zn_{0.05}Te:Cl$ crystals, the conduction inversion $p \longrightarrow n$ is observed at higher carrier densities, $p(n) \geq 10^9$ cm$^{-3}$. The curve describing the $n(P_{CA})$ dependence for the CdTe:Cl crystals is very steep: $n$ increases from $10^7$ to $10^{15}$ cm$^{-3}$ under only an approximately twofold increase in the partial pressure $P_{Cd}$. For the $Cd_{0.95}Zn_{0.05}Te:Cl$ crystals, the curve exhibits a more gentle slope: when $P_{Cd}$ increases 10 times, the carrier density $n$ increases from $10^7$ to $10^{11}$ cm$^{-3}$ (while the material remains virtually semi-insulating). In contrast to the case of CdTe:Cl, this behavior of the $n(P_{Cd})$ dependence in the $Cd_{0.95}Zn_{0.05}Te:Cl$ crystals creates considerably more favorable conditions for controlling the electrical properties of a crystal during its fabrication, specifically, by varying the partial pressure of the Cd vapor.

The low free-carrier density, $p(n) = 10^7$-$10^8$ cm$^{-3}$, in the $Cd_{1-x}Zn_xTe:Cl$ crystals is related to self-compensation of charged point defects. According to one of the models of self-compensation [9–11], a strong relaxation of the crystal lattice in the vicinity of Cl atoms leads to the formation of $DX^-$ centers, which generates deep levels below the minimum of the conduction band. This circumstance accounts for the formation of semi-insulating $n$-$Cd_{1-x}Zn_xTe:Cl$ crystals. However, two points remain unaccounted for:

(i) The dependence of the electron density on $P_{Cd}$; indeed, in terms of this model, the lattice relaxation depends only on the composition of the solid solution (Zn content) and the nature of the dopant (Cl). (ii) The formation of $DX^-$ centers in $Cd_{1-x}Zn_xTe:Cl$ becomes energetically favorable only at the Zn content $x \geq 0.2$ [10]. It is necessary to note that the formation of $DX^-$ centers has been observed in $Cd_{1-x}Zn_xTe:Cl$ crystals with $x < 0.2$ [9, 11].

In terms of the other model [3–8], the self-compensation is accounted for by the formation of intrinsic point defects, i.e., metal vacancies ($V_{Zn}$ and $V_{Cd}$), which act as acceptors, and by the association of charged donors and acceptors into centers that give rise to deep levels in the band gap. According to this model, the concentration of $V_{Cd}$ acceptors in the crystal is high at low $P_{Cd}$. This circumstance leads to a high density of free holes, which can be seen in Fig. 1a for the CdTe:Cl and $Cd_{0.95}Zn_{0.05}Te:Cl$ crystals. As the pressure of the Cd vapor on the crystal increases, the concentration of $V_{Cd}$ decreases. The dependence of $p$ on $P_{Cd}$ for the CdTe samples demonstrates a corresponding decrease in the hole density. For the $Cd_{0.95}Zn_{0.05}Te:Cl$ samples, we did not observe any clearly pronounced $p(P_{cd})$ dependence. The high hole density and the lack of a clear $p(P_{Cd})$ dependence indicate that the $V_{Cd}$ concentration remains high. The high value of $[V_{Cd}]$ is a consequence of insufficiently high $P_{Cd}$ on the $Cd_{0.95}Zn_{0.05}Te:Cl$ crystal during the annealing. (Indeed, the partial pressure $P_{cd}$ for stoichiometric $Cd_{0.95}Zn_{0.05}Te:Cl$ must be higher than for CdTe at the same temperature [17].) At higher $P_{Cd}$, the concentration of donor centers, $[Cd_I] + [Cl_{Te}]$, exceeds the $[V_{Cd}]$ concentration. This behavior of the concentrations of charged defects correlates with the data in Fig. 1a, in which the conduction inversion $p \longrightarrow n$ is observed for CdTe:Cl and $Cd_{0.95}Zn_{0.05}Te:Cl$. For CdTe:Cl, the $p \longrightarrow n$ inversion is sharp (almost step-like). This variation in the carrier density is related to a "jump" in the Fermi level from a deep acceptor to a shallow donor ($Cd_I$ and $Cl_{Te}$) level.

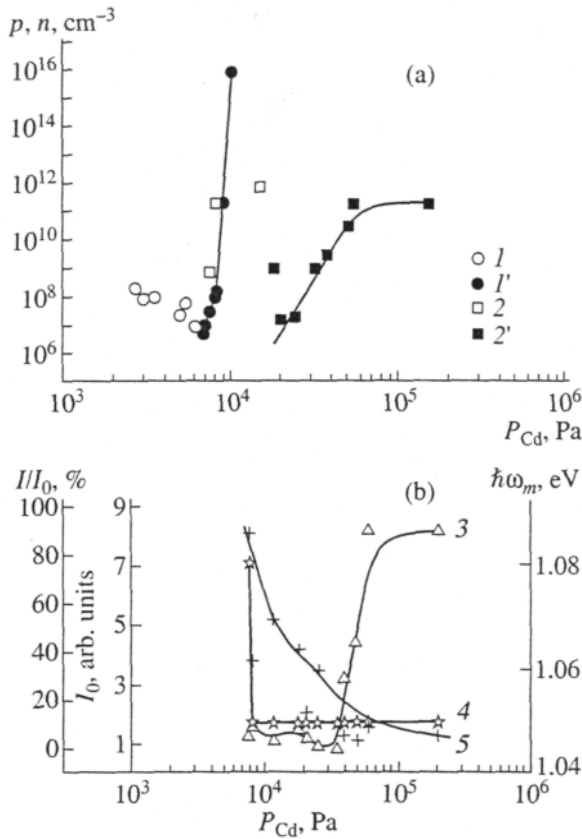

Fig. 1. (a) Carrier density and (b) PL characteristics as func-tions of the Cd vapor pressure during annealing of the sam-ples, (a): (1) $p$-CdTe, (1') $n$-CdTe, (2) $p$-$Cd_{0.95}Zn_{0.05}Te$, and (2') $n$-$Cd_{0.95}Zn_{0.05}Te$. (b): (3) relative integral intensity $I/I_0$ of the luminescence band at 1 eV, (4) the position of peak of the 1-eV band $\hbar\omega_m$, and (5) total integral PL intensity $I_0$.

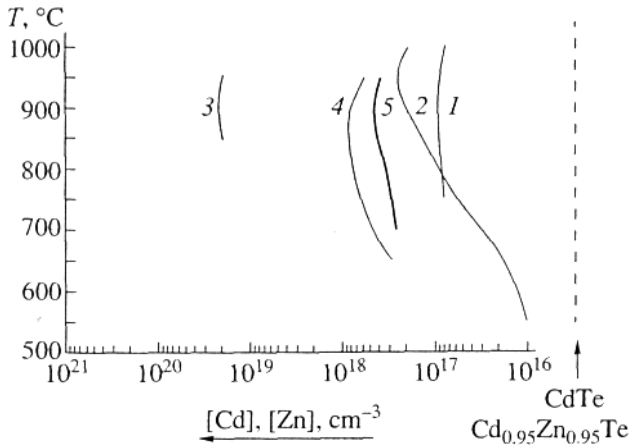

Fig. 2. Maximum solubility of Cd in CdTe: (1, [3]; 2, [20]; 3, [18]; and 4, [19]) and of Cd and Zn in $Cd_{0.95}Zn_{0.05}Te$ (5, [17]) along the $T$-$x$ section of the state diagram of the Cd-Zn-Te system.

The rate at which the electron density $n$ grows with $P_{Cd}$ is considerably lower in the $Cd_{0.95}Z_{0.05}Te{:}Cl$ crystals than for CdTe:Cl. This fact may indicate that there is a weak dependence of the concentration of donor point defects $[Cd_I]$ and $[V_{Te}]$ on $P_{cd}$. The limited solubility of these intrinsic point defects in $Cd_{1-x}Zn_xTe{:}Cl$ can be seen in the state diagram of the Cd-Zn-Te system shown in Fig. 2. The limit of the maximum Cd concentration in a $Cd_{0.95}Z_{0.05}Te$ crystal (curve 5) indicates that the $Cd_I$ and $V_{Te}$ solubility is lower than that in CdTe (curves 3, 4). These curves reflect the total concentration of point defects. Furthermore, the limited $V_{Te}$ solubility in the crystal results in limited solubility of the principal donor dopant substituting Te, i.e., $Cl_{Te}$. Therefore, from the data on the solubility of $V_{Te}$ in $Cd_{0.95}Zn_{0.05}Te$ and CdTe, we may assume a lower solubility of Cl in the solid solution as compared to the binary compound. Prior to annealing, Cl in the $Cd_{0.95}Z_{0.05}Te{:}Cl$ samples exists in charged state at the concentration $[Cl^+_{Te}] \sim 10^{17}$ cm$^{-3}$, which is necessary to obtain a semi-insulating material with $p \sim 10^8$ cm$^{-3}$. During annealing at 900°C, the concentration of $Cl^+_{Te}$ defects can decrease due to egress of Cl to some inclusions in the crystal (or to a gaseous phase) or as a result of the retrograde character of the solubility of these defects in the crystal. As the content of $Cl^+_{Te}$ and $Cd^+_I$ decreases, the concentration of centers forming shallow levels in the band gap also decreases; therefore, the transition of the Fermi level from deep to shallow levels becomes more gradual. These considerations furnish a qualitative explanation of the difference between the dependences of the electron density on $P_{Cd}$ in $Cd_{0.95}Zn_{0.05}Te$ and CdTe (Fig. 1a).

The quantitative estimate is hindered by the following:

(i) The available experimental data on the solubility of point defects, which were obtained by different authors using different methods, strongly diverge from each other and from the results of calculations (see Fig. 2).

(ii) Until now, the concentration of charged point defects ($V_{Zn}$ and $Zn_I$) in $Cd_{0.95}Zn_{0.05}Te$ has not been studied.

iii) The solubility of Cl in $Cdg_{0.95}Zn_{0.05}Te$ has not yet been determined.

We have tried to estimate the maximum solubility of charged intrinsic donor defects using recent data on the total solubility of defects in $Cd_{0.95}Zn_{0.05}Te$ [17] and CdTe [19] and the data on charged intrinsic point defects in CdTe [3, 20] (the last data show only a slight scatter, see curves 1 and 2 in Fig. 2). We assume that the ratio between the total amounts of intrinsic point defects and charged defects in $Cd_{0.95}Zn_{0.05}Te$ is the same as in CdTe:

$$\left(\frac{[Cd_i^+] + [V_{Te}^+]}{[Cd_i] + [V_{Te}]}\right)_{CdTe} = \left(\frac{[Cd_i^+] + [Zn_i^+] + [V_{Te}^+]}{[Cd_i] + [Zn_i] + [V_{Te}]}\right)_{Cd_{0.95}Zn_{0.05}Te}. \quad (1)$$

This assumption is based on the fact that the crystal lattice energies in CdTe and $Cd_{0.95}Zn_{0.05}Te$ differ insignificantly [21] and the samples are annealed in similar $P$-$T$ (pressure-temperature) conditions. We also assume that the Cd vapor pressure governs change in the concentration of charged defects (the Fermi level position and free-carrier density). The data on the concentration of defects at 900°C are taken from Fig. 2. For $Cd_{0.95}Zn_{0.05}Te$, we obtain $[Cd^+_I] + [Zn^+_I] + [V^+_{Te}] = 5 \times 10^{16}$ cm$^{-3}$. The electron density in the conduction band must be the same at high $P_{Cd}$; however, this is not the case (see Fig. 1a). Therefore, either assumption (1) is erroneous or the used concentration of point defects at 900°C does not correspond to the concentration of defects in the semi-insulating crystal at room temperature, at which the Hall effect was measured. If we assume that assumption (1) is correct and use it in the estimation of the concentration of defects at low temperatures (~500°C), when the equilibrium of the defects is "frozen" during the cooling of the crystal, we obtain an even higher value of $[Cd^+_I] + [Zn^+_I] + [V^+_{Te}]$, since the solubility of the metal in $Cd_{0.95}Zn_{0.05}Te$ at low temperatures is higher than in CdTe [17]. It is necessary to note, however, that, while discussing the dependence of the ratio on $P_{Cd}$, we have formally included $[Zn^+_I]$ in it, and we expect a conduction inversion through variation of the ratio $([V^-_{Cd}]+[V^{2-}_{Cd}])/[Cd^+_I]$ with $P_{Cd}$.

$$\left(\frac{[Cd_i^+] + [Zn_i^+] + [V_{Te}^+]}{[Cd_i] + [Zn_i] + [V_{Te}]}\right)_{Cd_{0.95}Zn_{0.05}Te} \quad (2)$$

Tab. 1: Product of the carrier mobility by lifetime for electrons and holes and the photocurrent memory in samples annealed at different $P_{Cd}$

| Before annealing | | | After annealing | | | | |
|---|---|---|---|---|---|---|---|
| $(\mu\tau)_e$, cm$^2$ V$^{-1}$ | $(\mu\tau)_h$, cm$^2$ V$^{-1}$ | photocurrent memory, % | annealing no. | $P_{Cd}$, $10^4$ Pa | $(\mu\tau)_e$, cm$^2$ V$^{-1}$ | $(\mu\tau)_h$, cm$^2$ V$^{-1}$ | photocurrent memory, % |
| ~$2 \times 10^{-4}$ | ~$(1-3) \times 10^{-5}$ | 9–10 | 063 | 1.8 | $5 \times 10^{-5}$ | $4.5 \times 10^{-5}$ | 3.5 |
| | | | 064 | 0.8 | $4 \times 10^{-5}$ | $<2 \times 10^{-7}$ | 0.13 |
| | | | 067 | 2.5 | $5 \times 10^{-5}$ | $<8 \times 10^{-7}$ | 0.4 |
| | | | | | $3.5 \times 10^{-5}$ | $<2 \times 10^{-6}$ | |
| | | | | | $6 \times 10^{-5}$ | | |
| | | | 070 | 5 | ~$2.5 \times 10^{-7}$ | $<5 \times 10^{-7}$ | 0.17 |
| | | | | | ~$2.5 \times 10^{-6}$ | $<2 \times 10^{-7}$ | |

The concentration of point defects involving Zn in a sample is con-trolled by the composition and temperature of the filling material in the ampule, and these parameters for the filling and the sample coincide. The hole density $p = 10^8\text{-}10^9$ cm$^{-3}$, which was measured in the samples before annealing, indicates that the intrinsic point defects with the greatest concentrations are [V$^-_{Cd}$] and [V$^{-2}_{Cd}$], and [V$^-_{Zn}$] and [V$^{-2}_{Zn}$]. During annealing at high $P_{Cd}$, the concentration of [V$^-_{Cd}$] and [V$^{-2}_{Cd}$] in the metal sublattice decreases, whereas that of [Cd$^+_I$] and [V$^+_{Te}$] increases. In this situation, the concentrations of intrinsic point defects [V$^-_{Zn}$] and [V$^{-2}_{Zn}$], which are not directly related to $P_{cd}$, may remain stable or change only slightly. The presence of a deep level in the band gap, associated with a center related to a Zn vacancy, can account for the slow increase in $n$ with $P_{Cd}$ occur-ring under a sharp increase in the concentration of shallow donors [Cd$^+_I$] and [Cl$^+_{Te}$].

Along with the Hall effect measurements, the photoluminescence (PL) spectra, carrier lifetime and mobility, and the photocurrent memory were measured in all the Cd$_{0.95}$Zn$_{0.05}$Te samples annealed at various pressures $P_{Cd}$.

PL was measured at 77 K in the range of photon energies $\hbar\omega_m$ = 0.8-2.3 eV. The emission was excited by an Ar-ion laser (the photon energy was 2.43 eV and the flux density was ~$10^{20}$ cm$^{-2}$ s$^{-1}$) and detected by a cooled Ge photodiode. As is standard, the PL spectrum consisted of three bands peaked at about 1.6, 1.45, and 1 eV. We now restrict our consideration to the 1-eV band. Figure Ib shows the ratio of the integral intensity of this band $I$ to the total intensity $I_0$ integrated over the whole spectrum (curve *3*), the peak position $\hbar\omega_m$ of the band (curve *4*), and the total integral intensity $I_0$ (curve *5*) as functions of $P_{Cd}$. The integral intensities were measured under the same excitation and detection conditions for different samples. It can be seen that $I/I_0$ remains virtually unchanged, at about 6%, up to $P_{Cd} = 3.5 \times 10^4$ Pa. It then increases to -90% and tends to a constant value. It is noteworthy that the curves describing the electron density $n$ and of the intensity $I$ of the PL band level out in the same range of $P_{Cd}$ (in Fig. 1a, curves 2', Fig. 1b, curve 3). The fact that the spectral position of the band peak remains constant (at least for $P_{Cd}$ corresponding to the *n*-type samples) indicates that the increase in the band intensity is related to an increase in the concentration of centers of the same kind rather than to the appearance of some other type of centers. The origin of the 1-eV band still remains unclear. Presumably, it has a complex nature [22, 23]. Several authors, including us, have attributed it to the capture of carriers by centers that are either isolated doubly negatively charged Cd vacancies, V$^{-2}_{Cd}$ [3], or various complexes that include these vacancies [15, 24-27]. Under the conditions of our experiment, the concentration of Cd vacancies cannot increase with the pressure of the Cd vapor. However, it seems quite probable that the initial sample contains V$_{Zn}$ vacancies, and, as was mentioned above, they are retained under annealing with a filling. This assumption is supported by the estimate of the energy position of the V$_{Zn}$ level in ZnTe [23], which nearly coincides with the $V_{Cd}$ level in CdTe. Under annealing at higher $P_{Cd}$, the concentration [V$_{Me}$D] of acceptor complexes constituted by a metal vacancy and a donor, which are responsible for the PL band at 1.45 eV, decreases, with the result that the contribution of the 1-eV band increases.

As can be seen in Fig. Ib, an increase in the Cd vapor pressure during annealing reduces the integral PL intensity $I_0$. This observation indicates an increase in the concentration of nonradiative recombination centers (or emission outside the spectral range under study, which seems unlikely). This dependence correlates with the data obtained for the product of the carrier mobility and lifetime and for the photocurrent memory in the samples annealed at high $P_{Cd}$. These data are presented below.

Using the time-of-flight method and irradiation with a particles (see, e.g., [28]), we measured the product of the carrier mobility and the lifetime for electrons $(\mu\tau)_n$ and holes $(\mu\tau)_h$ in the crystals under study before and after annealing (see table). It can be seen that the annealing changes $(\mu\tau)_e$ only slightly. At the same time, $(\mu\tau)_h$ in the annealed crystals is significantly reduced. This fact confirms the above-made assumption that the annealing of Cd$_{0.95}$Zn$_{0.05}$Te crystals at high $P_{Cd}$ leads to the formation of deep acceptor centers, which are possibly the intrinsic point defects V$^-_{Zn}$ and V$^{-2}_{Zn}$. These centers are negatively charged, which results in an intense capture of holes.

The study of the same crystals before and after annealing by intense pulses of uniformly absorbed radiation (the computer tomography mode [29, 30]) has demonstrated the strong photocurrent memory present in the unannealed samples (see table). It can be seen that, after annealing at low $P_{Cd}$ close to the pressure at which the $p \rightarrow n$ inversion occurs (annealing no. 063), the photocurrent memory is slightly reduced. In annealing nos. 064, 067, and 070 at higher $P_{Cd}$, the photocurrent memory is strongly reduced: to (0.13-0.4)%. The magnitude of the photocurrent memory correlates with the dependence of $(\mu\tau)_h$ on $P_{Cd}$ (see table). As is known [15,29, 30], one of the origins of photocurrent memory is an incomplete collection of the hole component of the signal. In our case (annealing nos. 064, 067, 070), the hole component is virtually absent, which results in a strong decrease (by a factor of 10-20) in the photo-current memory and indicates that there appears a large amount of deep acceptor levels responsible for the recombination of holes.

Thus, study of the self-compensation in single-crystal $Cd_{0.95}Zn_{0.05}Te:Cl$ shows that the $p \rightarrow n$ conduction inversion is observed at a higher free-carrier density ($n, p \sim 10^9$ cm$^{-3}$) than in CdTe:Cl crystals. In addition it was found that the curve describing the dependence of the electron density on the Cd vapor pressure exhibits a more gentle slope.

Studies of photoluminescence, carrier lifetime and mobility, and photocurrent memory have confirmed the formation of a deep acceptor level in the band gap of the annealed crystals. This finding indicates that the low conductivity of annealed $Cd_{0.95}Zn_{0.05}Te:Cl$ crystals can be explained in terms of a model of self-compensation via the formation of charged intrinsic point defects with a deep acceptor level. We assume that this level is related to Zn vacancies, which remain active at high $P_{Cd}$.

It is noteworthy that the revealed dependence of the electron density in $Cd_{0.95}Zn_{0.05}Te:Cl$ on $P_{Cd}$ creates considerably more favorable conditions for controlling the electrical properties of a crystal by varying the Cd vapor pressure during the fabrication of the material in comparison with CdTe:Cl.

The study was supported by INTAS (project no. 99-1456).